\documentclass[reprint,showpacs,amsmath,amssymb,aps,prb,superscriptaddress]{revtex4-1}

\usepackage{graphicx}
\usepackage{hyperref}

\def\Tc{\ensuremath{T_\text{c}}}
\def\Hc{\ensuremath{H_\text{c2}}}
\def\cel{\ensuremath{c_\text{el}}}

\def\LiPdPt{Li$_\text{2}$(Pd$_{\text{1}-x}$Pt$_x$)$_\text{3}$B}
\def\acchi{\ensuremath{\chi_\text{ac}}}
\def\HT{\ensuremath{H}--\ensuremath{T}}

\begin{document}

\title{Magnetic Phase Diagram of \LiPdPt\ by AC Susceptometry}

\author{D. C.~Peets}
\email{dpeets@fkf.mpg.de}
\altaffiliation[Current address:  ]{Max Planck Institute for Solid State 
Physics, Heisenbergstr. 1, D-70569 Stuttgart, Germany}
\affiliation{Department of Physics, Graduate School of Science, Kyoto
University, Kyoto 606-8502, Japan}

\author{G.~Eguchi}
\affiliation{Department of Physics, Graduate School of Science, Kyoto
University, Kyoto 606-8502, Japan}

\author{M.~Kriener}
\altaffiliation[Current address:  ]{Institute of Scientific and Industrial 
Research, Osaka University, Osaka 567-0047, Japan}
\affiliation{Department of Physics, Graduate School of Science, Kyoto
University, Kyoto 606-8502, Japan}

\author{S.~Harada}
\author{Sk.\ Md.\ Shamsuzzamen}
\affiliation{Department of Physics, Okayama University, Okayama 700-8530, Japan}

\author{Y.\ Inada}
\affiliation{Faculty of Education, Okayama University, Okayama 700-8530, Japan}

\author{G.-Q.\ Zheng}
\affiliation{Department of Physics, Okayama University, Okayama 700-8530, Japan}

\author{Y.~Maeno}
\affiliation{Department of Physics, Graduate School of Science, Kyoto
University, Kyoto 606-8502, Japan}

\date{\today}

\begin{abstract}

The \HT\ phase diagram and several superconducting parameters for \LiPdPt\ have been determined as a function of cation substitution $x$.  Notably, the coherence length may be linear in platinum concentration.  Despite the superconducting pairing state and band structure apparently changing fundamentally, the \HT\ phase diagram is essentially unchanged.  Unusual aspects of the shape of the \HT\ phase diagram are discussed.  The upper critical field \Hc(0) is not anomalously high for any Pt content, likely due to an absence of high carrier masses --- in such a case, the value of \Hc(0) would not serve as a probe for novel physics.  

\end{abstract}


\maketitle

\section{Introduction}\label{sec:intro}

Most superconductors have crystal structures containing an inversion center, such that transposing every atomic position through that point returns the original crystal structure.  In such a superconductor, parity constrains the superconducting pairing state to be either singlet or triplet, and the basic physical properties of this state, if not always the underlying mechanisms, are well-known.  However, noncentrosymmetric superconductors, those which lack spatial inversion, also exist, and such materials can exhibit extremely unconventional behavior.  The first indication of novel physics in such a system was in CePt$_3$Si, \cite{Bauer2004} which was found to have an extremely high, non-Pauli-limited upper critical field \Hc.  This touched off strong interest in noncentrosymmetric materials, with systems as diverse as UIr,~\cite{Akazawa2004} B-doped SiC,~\cite{Ren2007} Mo$_3$Al$_2$C,~\cite{Bauer2010} Mg$_{10.5}$Ir$_{19}$B$_{17.1}$,~\cite{Klimczuk2006} and 1-1-3 silicides such as CeRhSi$_3$,~\cite{Kimura2005} BaPtSi$_3$,~\cite{Bauer2009} or CaIrSi$_3$~\cite{Eguchi2010,Eguchi2011} being studied.  A non-Pauli-limited \Hc\ remains a key signature of novel behavior in noncentrosymmetric superconductors.

In the absence of inversion symmetry, parity is no longer a meaningful concept, and the pairing state can thus no longer be strictly classified as singlet or triplet.  Spin-orbit terms that otherwise cannot contribute due to symmetry can split the band structure and Fermi surface by spin orientation.  This can lead to a spin imbalance at the Fermi surface and, if the splitting is large compared to the superconducting gap, to pairing that is largely constrainted to be within each Fermi-surface sheet, implying an admixture of singlet and triplet components.~\cite{Frigeri2004,Fujimoto2007,Yanase2007}  While much theoretical work remains to be done on these materials, a remarkable array of exotic phases and responses have been predicted, from unusual magnetoelectric effects~\cite{Edelstein1995} to FFLO-like pair density wave states with spatially-varying pairing functions, states with helical nodes around vortices, or states with half- or quarter-quantum vortices, in the presence of a small magnetic field.~\cite{Agterberg2007,Agterberg2008PRL,Agterberg2008NPhys,Matsunaga2008}

Perhaps since most of the exotic effects arising from the lack of spatial inversion require that the bands exhibit significant spin-orbit splitting at the Fermi surface, the majority of the noncentrosymmetric systems studied to date have been found to behave in a fairly conventional manner.  Indeed, unusual behavior has been reported almost exclusively in heavy-electron lanthanide and actinide systems, and a great deal of the predicted physics remains unobserved.  By far the clearest evidence for unconventional behavior without heavy electrons is in \LiPdPt, which appears to be fully gapped and singlet-dominated for $x=0$ but exhibits line nodes and strong triplet character for $x=1$,~\cite{Yuan2006,Nishiyama2007} with indications of unconventional behavior as early as $x=0.2$.~\cite{Badica2010}  Given this rather fundamental change in behavior with cation substitution, and noting that the most unusual consequences stemming from the lack of inversion are manifested in the magnetic properties, a detailed magnetic phase diagram for the substitution series would be of interest.  Indeed, the shape of the upper critical field as a function of temperature may indicate the dominant pairing symmetry and the strengths of various interactions in these materials, and several predictions have been made for the \HT\ phase diagram.~\cite{Agterberg2007,Yanase2007,Samokhin2008}  While portions of an \HT\ phase diagram have been reported for several substitution levels based on $M$--$H$ loops,~\cite{Badica2005} it was not possible to resolve the shape or especially the low-temperature behavior, and a lack of low-temperature points made extrapolation to a reliable \Hc(0) difficult.  The primary purpose of this paper is to provide the full substitution dependence of the magnetic phase diagram.

\section{Experimental}\label{sec:expt}

Polycrystalline ingots of \LiPdPt\ were prepared from Li (99.9\% purity), Pt (99.999\%), Pd (99.95\%), and B (99.8\%) by a two-step arc melting process, with a slight excess of lithium used to compensate for losses due to evaporation.~\cite{Togano2004}  Pieces mechanically separated from the ingot were measured by $ac$ magnetic susceptometry by a mutual-inductance technique in a homebuilt first-derivative coil, mounted in a $^3$He refrigerator (Oxford Instruments) inserted into a 9~T magnet.  Measurements were performed at frequencies of around 800~Hz, using an $ac$ magnetic field of 100~$\mu$T $rms$ parallel to the dc field, and measured using a lock-in detector.  The temperature or field was swept over the course of several hours.  No hysteresis in temperature was observed that could not be attributed to rapid cooling, so all data presented were measured on warming, for which the temperature could be changed more controllably.  It was necessary to subtract a field-dependent background signal, precluding analysis of the normal-state susceptibility.  The upper critical field was defined as the point at which the diamagnetic signal $\acchi'$ reached 5\% of its full, zero-field, low-temperature value.  

For the purposes of comparison, the specific heat $c_P$ was also measured on several samples, by a thermal relaxation method using a commercial calorimeter on a $^3$He refrigerator (Quantum Design, PPMS), between 0.3 and 30 K.  Measurements were taken with and without each sample present, to isolate the effect due to the sample.  Data collected on warming and cooling were consistent.  The electronic specific heat \cel\ was separated from the phononic contribution by fitting data taken above \Hc(0) and subtracting $T^3$ and higher-order contributions.  The full specific heat study will be described elsewhere.  In the specific heat data, \Hc\ was defined as the midpoint of the superconducting transition.

\section{Results and Analysis}

\begin{figure}[htb]
\includegraphics[width=\columnwidth]{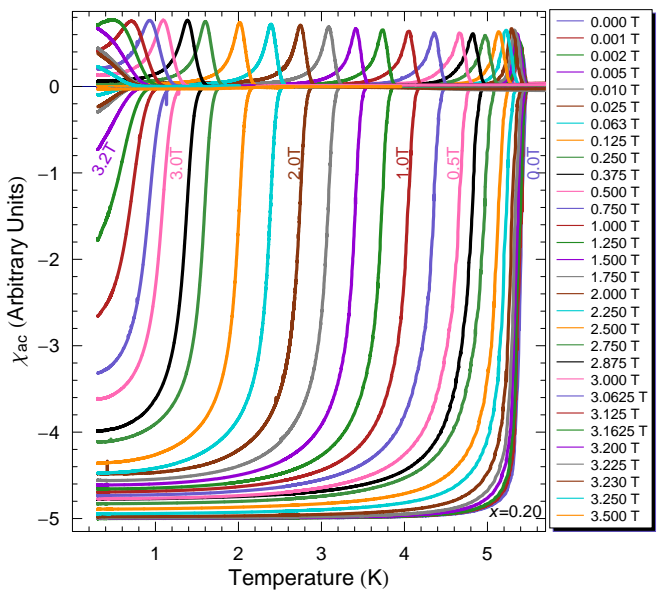}
\caption{\label{fig:20percent}(color online) Temperature and field dependence of the diamagnetic shielding $\acchi'$ (negative) and loss $\acchi''$ (positive) for \LiPdPt\ with $x = 0.20$.
}
\end{figure}

Figure \ref{fig:20percent} shows the temperature dependence of the ac susceptibility of a \LiPdPt\ sample with $x=0.20$.  The narrow transitions indicate good homogeneity and sample quality.  After some initial broadening in very low fields, the width changes little until very close to \Hc(0), at which point the transitions broaden rapidly.  The decrease in transition temperature with field appears to be fairly constant over much of the phase diagram.  No unusual glitches or secondary transitions are observed that would suggest the presence of additional phase transitions, although small peaks just below the superconducting transition, attributed to depinning of the vortex lattice, were observed in some samples when they were warmed after rapid cooling (not shown).

\begin{figure}[htb]
\includegraphics*[width=\columnwidth]{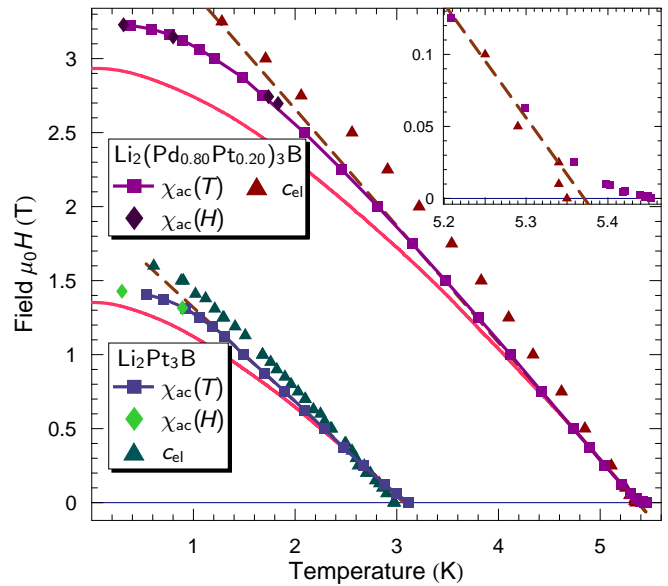}
\caption{\label{fig:withCel}(color online) Example of a resulting \Hc\ curve for the Pt endmember ($x=1$) and for $x=0.20$, including \cel\ points taken on the same $x=1$ sample and a larger $x=0.20$ from the same ingot --- the difference in slope may be due to having the sample in a different orientation relative to the field, and should not be taken as strong evidence for a vortex melting transition.  The curve predicted by dirty-limit WHH (with the \Tc\ and initial slope taken from the linear fit) is included for comparison, and the inset shows the strong upward curvature detected in \acchi\ near \Tc, for $x=0.20$.
}
\end{figure}

Figure \ref{fig:withCel} shows example \HT\ phase diagrams derived from such data, for $x=1$ (the Pt endmember) and $x=0.20$.  The upper critical field is indeed remarkably linear over much of the phase diagram, then turns over rapidly at high fields.  Transitions extracted from specific heat data are included for comparison --- the shape is similar, but the slope is higher.  The near-agreement confirms that the measured $\acchi$ is bulk-sensitive.  Areas where the \acchi\ data fall below the \cel\ points could indicate a vortex liquid state, in which field penetrates the sample almost completely and the vortices are mobile, mimicking normal-state response to ac stimulus.  However, since the polycrystalline samples had irregular shapes, their microstructure is unknown, and their orientation was most likely not the same in both experiments where the same sample was used, it is also quite possible that this difference is attributable to grain structure or demagnetizing effects.  It should also be noted that, in the Pt endmember at least, pinning has been found to be extraordinarily strong.~\cite{Miclea2009}

Despite the unconventional behavior observed in the Pt endmember by other techniques, the high, non-Pauli-limited \Hc\ generally associated with noncentrosymmetric physics is not observed.  Preliminary specific heat data on the samples studied here and previous reports in this system~\cite{Togano2004,Takeya2007,Hafliger2009} all indicate a normal-state electronic specific heat coefficient $\gamma_n$ of about 7-10~mJmol$^{-1}$K$^{-2}$, consistent with ordinary $d$-electron metals.  In order to see Pauli depairing, it is necessary that orbital effects not break the pairs first.  In systems such as CePt$_3$Si, heavy Fermion masses lead to a suppression of orbital pair-breaking effects, and the absence of Pauli pair breaking is clear.  In \LiPdPt, however, orbital limiting is not suppressed, and \Hc\ may not probe Pauli limiting effects.  Thus, while no anomalously high \Hc\ is observed, this should not be taken as evidence for an absence of novel physics.  

The shape predicted for $\Hc(T)$ in the WHH dirty limit calculations,~\cite{Helfand1966,Werthamer1966} based on BCS theory, is also shown in Fig.~\ref{fig:withCel} for each sample.  This curve is most commonly used to predict an approximate value of \Hc(0) from \Tc\ and $d\Hc/dT|_{T=\Tc}$ when the data do not allow extrapolation to zero field.  For the purposes of the comparison in Fig.~\ref{fig:withCel}, the main slope rather than the initial slope was used, due to curvature near \Tc which is in clear violation of WHH expectations (see inset for $x=0.20$) and may not represent intrinsic behavior.  Even neglecting the lowest fields, the WHH curve fails to fully capture the shape of $\Hc(T)$.  Within a conventional picture, this could possibly be understood in terms of a breakdown of one or more assumptions underpinning the WHH calculations, for instance by multi-band or strong-coupled superconductivity.  

A variety of predictions have been made for \HT\ phase diagrams in noncentrosymmetric systems,~\cite{Agterberg2007,Yanase2007,Samokhin2008} but these curves generally exhibit strong downward curvature near \Tc\ and do not capture the other essential features of the \HT\ phase diagrams reported here.  The general case in the presence of impurities, however, cannot be solved analytically, and the results are strongly dependent on fine details of the band structure, pairing symmetry, and intra- and interband coupling.~\cite{Samokhin2010}

\begin{figure}[htb]
\includegraphics[width=\columnwidth]{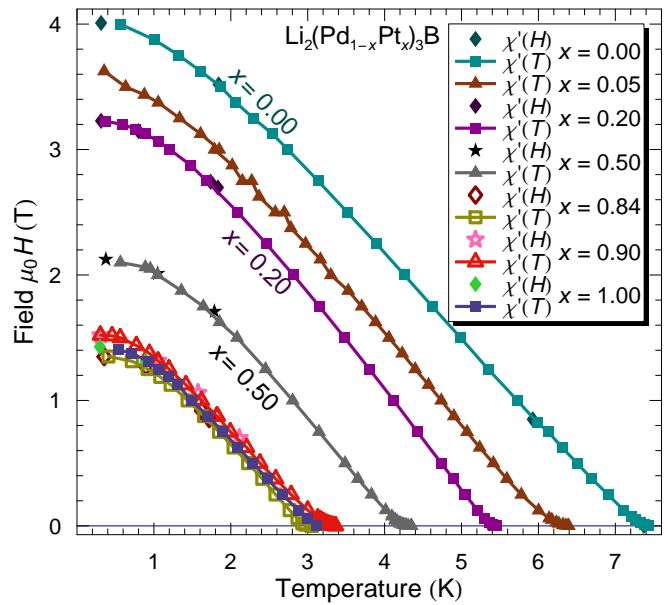}
\caption{\label{fig:combined}(color online) Combined phase diagram for \LiPdPt\ across the substitution range studied.  There is little variation in shape with cation substitution.
}
\end{figure}

The \HT\ phase diagram is shown as a function of cation substitution in Fig.~\ref{fig:combined}.  Since the Pt and Pd endmembers have very different band structure,~\cite{Lee2005} and since replacing Pd by Pt introduces nodes in the gap \cite{Yuan2006} and changes the NMR Knight shift and $1/T_1$ from singlet- to triplet-like behavior,~\cite{Nishiyama2005,Nishiyama2007} one might expect significant changes in the \HT\ phase diagram.  This is not observed.  The shape of the \HT\ phase diagram changes little with Pt content.  All curves exhibit upward curvature near \Tc, are remarkably linear over much of the field range, then curve over rapidly at high fields.  

\begin{figure}[htb]
\includegraphics[width=\columnwidth]{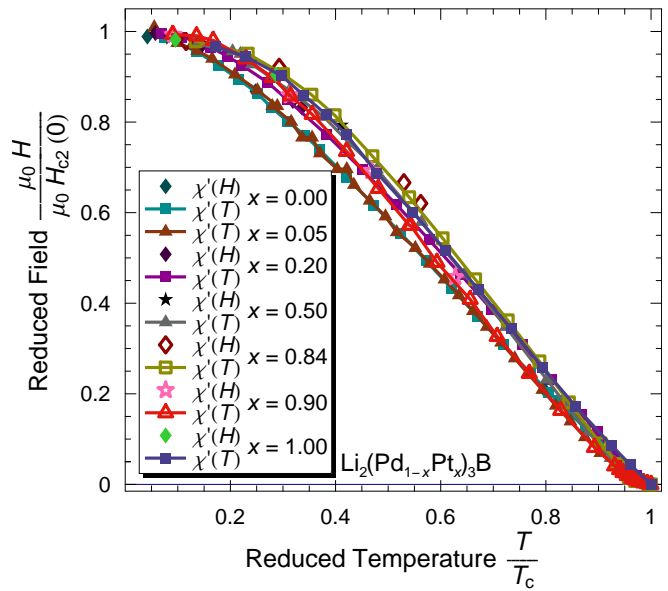}
\caption{\label{fig:scaling}\HT\ phase diagrams for all Pt concentrations studied, scaled by \Tc\ and \Hc(0).  There is no strong evidence for a systematic evolution with cation substitution.}
\end{figure}

To check for the presence of any evolution with cation substitution, Fig.~\ref{fig:scaling} plots all the \HT\ phase diagrams together, scaled by \Tc\ and \Hc(0).  While the data for different Pt contents do not collapse perfectly onto a single curve, no systematic evolution can be discerned.  Given the fundamental changes in the nature of the superconducting state and the differences in the band structure, this degree of uniformity from one end of the substitution phase diagram to the other is particularly striking.

\begin{figure}[htb]
\includegraphics[width=\columnwidth]{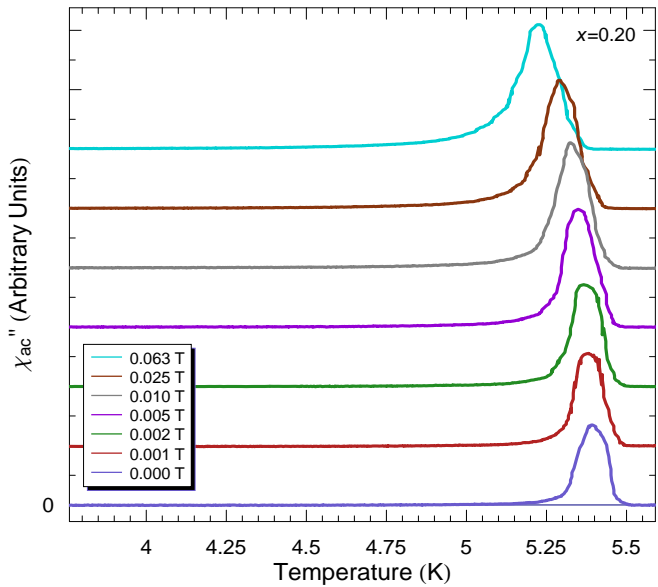}
\caption{\label{fig:waterfall}Transition in $\acchi''$ for very low dc fields, $x=0.20$ ($\mu_0\Hc(0) \approx 3.24$~T;  $H_\text{ac}=0.1$~mT $rms$).  The peak grows and broadens, developing a tail to low temperature.}
\end{figure}

It has been previously pointed out~\cite{Togano2004} that upward curvature in \Hc\ near \Tc\ as observed in this compound has also been seen in polycrystalline MgB$_2$ and borocarbides.~\cite{Takagi1994,Takano2001}  As very low fields are applied, the transitions broaden, an effect that is most pronounced in the $\acchi''$ fluctuation peak, the intensity of which also grows markedly (see Fig.~\ref{fig:waterfall}).  This suggests that grain boundary transport is strongly suppressed by field, and that the apparent upward curvature arises from critical current effects at junctions, rather than from entropy.  This may also offer an additional explanation for the small suppression of the magnetization-derived \HT\ phase diagram relative to that extracted from specific heat measurements --- \cel\ measurements are not sensitive to grain boundary transport.

\begin{figure}[htb]
\includegraphics[width=\columnwidth]{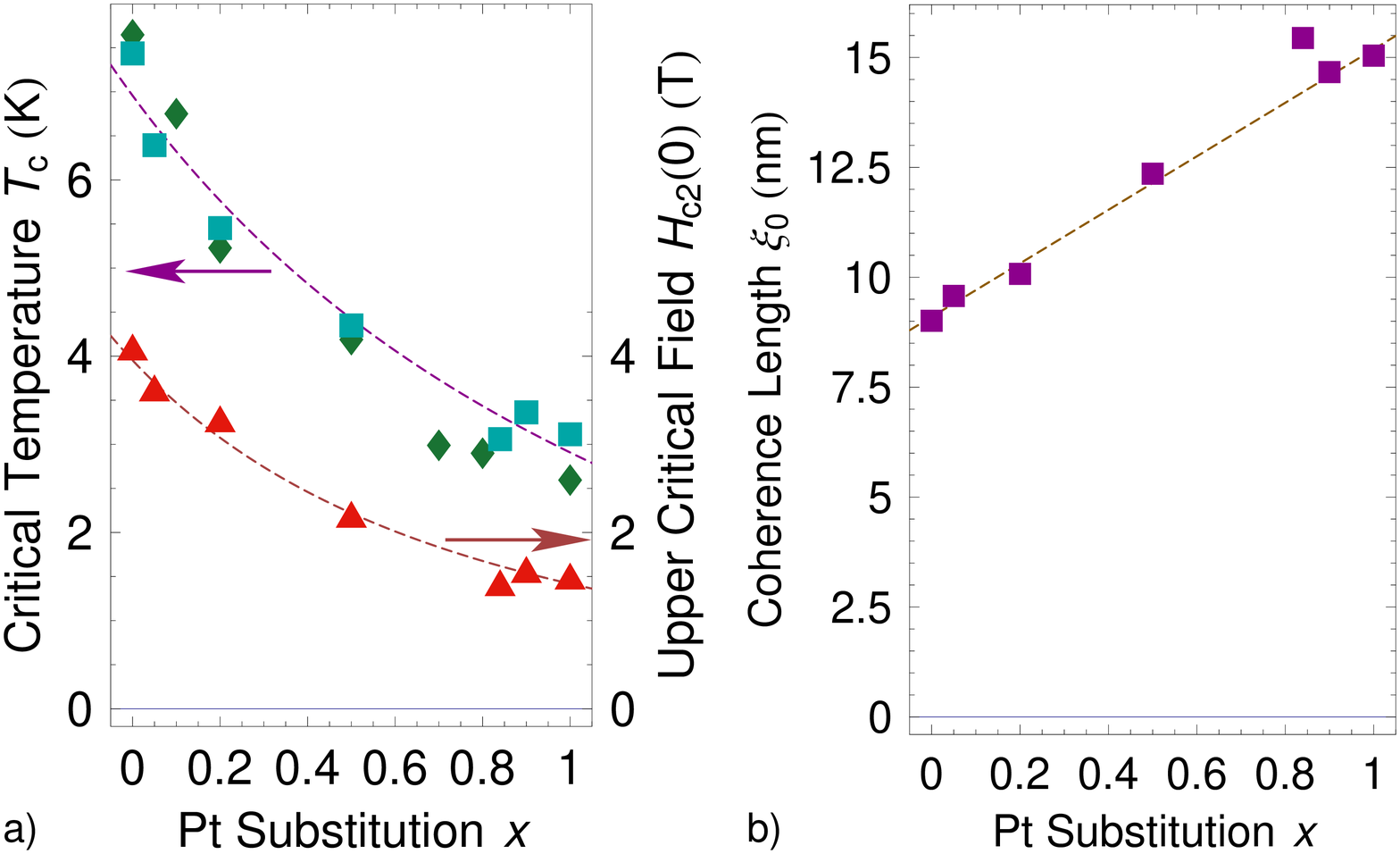}%
\caption{\label{fig:params}(color online) Dependence of \Tc, \Hc(0) and $\xi_0$ on cation substitution.  Diamonds are \Tc\ data from Ref.~\onlinecite{Badica2005}.
}
\end{figure}

The substitution-dependence of the transition temperature and the zero-temperature extrapolated upper critical field are presented in Fig.~\ref{fig:params}(a);  the superconducting coherence length calculated from \Hc(0) using the relation $\mu_0\Hc(0) = \Phi_0/2\pi \xi_0^2$ for orbital-limited fields is shown in Fig.~\ref{fig:params}(b).  The trend observed previously in \Tc\ tracks that previously reported,~\cite{Badica2005} but with significant deviations on the order of 0.5~K, perhaps due to the system being highly defect-sensitive or to a variable lithium concentration.  As might be expected given how little the \HT\ phase diagram changes with cation substitution, the substitution-dependence of \Hc(0) takes a similar shape.  Perhaps more surprisingly, the extracted zero-temperature coherence length appears to vary linearly with platinum content.  A coherence length linear in cation substitution would suggest $\Hc(0,x)\propto (x+a)^{-2}$ and, assuming the BCS $\xi_0(x) \propto \hbar v_{F}/k_B\Tc(x)$ to hold and $v_{F}(x)$ to be approximately constant, $\Tc\sim (x+a)^{-1}$.  The dashed lines in Fig.~\ref{fig:params} represent these power laws, with $a$ taken from $\xi_0$.  The significance of the coherence length varying linearly with cation substitution, if any, is not clear, and the possible sensitivity to defects in these materials requires that this result be treated with caution.

\section{Conclusion}

In summary, the \HT\ phase diagram of \LiPdPt\ has been determined as a function of $x$, and the substitution-dependence of several superconducting parameters has been elucidated --- most notably, the coherence length appears to vary linearly with platinum concentration.  Although the superconducting pairing state is thought to change fundamentally, from a fully-gapped, singlet-like to a triplet-dominated form with nodes, and the band structure changes substantially with cation substitution, the \HT\ phase diagram is essentially unchanged.  This striking result puts constraints on the changes seen in the pairing state, indicating a continuous evolution where an abrupt change would be expected.  Upward curvature near \Tc\ suggests poor transport at grain boundaries, while an abnormally linear, non-WHH shape indicates the superconducting state to be more complicated than simple single-band weak-coupling BCS.  The curve does not resemble predictions for noncentrosymmetic superconductors, but the shape expected in such a system can be extremely sensitive to a host of factors, including impurities, pairing symmetry, and fine details of the band structure.  The \Hc(0) values found in this study are not anomalously high as in CePt$_3$Si and some other materials, but this may be be attributable to an absence of the high carrier masses required to suppress orbital depairing and expose the presence or absence of Pauli limiting behavior.  In such a scenario, the value of \Hc(0) would not serve as an effective probe for the presence of novel physics.  

\begin{acknowledgments}
The authors thank S.\ Yonezawa for helpful discussions, and the Japan Society for the Promotion of Science; the Global COE program ``The Next Generation of Physics, Spun from Universality and Emergence'' from the Japanese Ministry of Education, Culture, Sports, Science, and Technology (MEXT); and the MEXT ``Topological Quantum Phenomena'' Grant-in Aid for Scientific Research on Innovative Areas.
\end{acknowledgments}

\bibliography{lithium}

\end{document}